# CALCULATION OF THE MINIMUM COMPUTATIONAL COMPLEXITY BASED ON INFORMATION ENTROPY

Xue Wu[1]

[1]Department of Microelectronics, Tsinghua University, Beijing, China

wuxiqh@yahoo.cn

## ABSTRACT

*In order to find out the limiting speed of solving a specific problem using computer, this essay provides a method based on information entropy. The relationship between the minimum computational complexity and information entropy change is illustrated. A few examples are served as evidence of such connection. Meanwhile some basic rules of modeling problems are established. Finally, the nature of solving problems with computer programs is disclosed to support this theory and a redefinition of information entropy in this filed is proposed. This will develop a new field of science.*

## 1. INTRODUCTION

As new approaches to algorithm optimization become more and more popular in researches, it is a matter of greater urgency to determine whether an algorithm of handling a specific problem reaches the limiting speed or not instead of trying to improve an algorithm continuously. Even though decision tree has been proposed to satisfy this demand, another technique is also established to discover the boundary of the velocity, based on information entropy.

Information is an abstract conception. There was not a widely acceptable measurement of information quantification until the father of information theory C. E. Shannon provided a novel conception of information entropy in 1948 [1], [2]. Shannon first associated probability with information redundancy in mathematic language [1]. His discovery made a great contribution to the field of communication, meanwhile it also left clues to the consistency between information entropy change and the nature of figuring out an issue using computer [3], [4], [5]. This directly led to the establishment of the first assumption.

Since computers were invented, the amount of information which is generated by an operation in a computer program has remained unknown to most people. The significance of operations' productivity has even been ignored. However it is necessary to concentrate on the efficiency of an operation with the purpose of building the bridge between information entropy change and the minimum computational complexity. Therefore the second assumption arises.

            

After the proposition of these two fundamental assumptions, there comes the great need of evidence to support this theory. Just then two examples appear and become the pillars of the theory. Looking for the maximum value and sorting a group of numbers are the names of these two problems. The results of the calculation based on information entropy for these two issues in some special models are the same with the computational complexities of the known fastest programs. This fact motivates a demand for notices of modeling these problems. With the help of the notices, the nature of solving problems using computer programs is discovered to testify the theory and a redefinition of the information entropy is proposed. This event heralds a new field of science to be developed [5].

## 2. TWO BASIC ASSUMPTIONS

### 2.1. Information Entropy Change and Solving a Problem

Considering some simple questions such as "Is it sunny today?" and "Is he a student?" answers to these problems usually make people know something that they don't understand before. In terms of this idea ravelling out some academic issues such as calculating the quantity of prime numbers and selecting the fastest network using computer indicates that there is a similar procedure with solving those simple problems.

During the information transmission and storage process, there is a part of code that does not express the substance of the information. The amount of such code is called information redundancy. Moreover making people realize some unknown knowledge is able to be regarded as giving people some substantial information [6]. Therefore the procedure is exactly the change of information redundancy. According to the relationship between information redundancy and information entropy, the first assumption that the nature of figuring out a problem is the same as a change of information entropy is set up.

### 2.2. The Amount of Information an Operation Generates

In the computer programs, problems are resolved by several operations. On the basis of the first assumption, settling a matter with the help of computers equals to changing the entropy of information. So operations in the programs have something to do with the amount of information. In other words, each operation produces a change in the information entropy.

No matter how complex an operation is, it is composed of some simple ones. Operations like equation, being greater than and being smaller than are typical instances of these basic operations. In consideration of them, each elementary operation is able to figure out a question like whether a statement is true or false. The amount of information that consists in such questions is regarded as 1 bit. Therefore the fact that every basic operation generates 1 bit information is presented as the second assumption.

## 3. THEORY DESCRIPTION MODELING OF TWO EXAMPLES

The fundamental role of information is to eliminate people's uncertainties of matters around them. The information entropy is used to measuring the degree of these uncertainties [6]. Shannon's conception of the entropy is based on the following equation:

$$H(x) = -\sum_{i=1}^{n} P(x_i) \log_2(P(x_i)) \qquad (1)$$

In the equation, H(x) is the information entropy, P(xi) is the probability of the incident that x equals to xi [7]. In line with the two assumptions a theory that the minimum computational complexity is the same with the change of information entropy is brought out. It can be expressed in the following form:

$$\text{Time complexity} = \Delta H(x) = H(x) - H_0(x) \quad (2)$$

H(x) is the initial entropy. H0(x) is the final entropy [8]. With regard to a specific question, the initial entropy is fetched through modeling of this problem [1]. The modeling criterion has not been clearly clarified, however an imprecise standard that the sequence which expresses the final relationships among data or belongs to the same sort of sequence with the final result should be regarded as the state xi is proposed. This rule will be interpreted by two examples in detail later on. When a problem is figured out, the information which people want to know is determined. In other words, the sequence is eventually decided. Therefore the probability of the situation that this sequence appears is 1 [9]. According to (1), the final entropy of an issue is obtained.

$$H_0(x) = \log_2(1)$$
$$= 0 \quad (3)$$

Backed by the modeling standard and the value of the ultimate information entropy, the minimum computational complexities of looking for the maximum value and sorting a group of numbers are easy to discuss.

### 3.1. Modeling of Seeking the Maximum Value

In order to make a model of looking for the maximum value, the form of the sequence that is able to expresses the characteristics of the final result should be decided first. We can abstract this problem into the following structure:

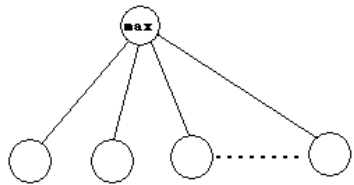

Figure 1. Abstract structure of seeking the maximum

The top element in Fig. 1 represents the maximum value and the other elements below it are the rest of the numbers. The lines between them are called keys. They denote the size relationships between the maximum value and the others. Each size relationship has two possible conditions: "Greater" or "Not greater". If each key's condition is determined as "Greater", then the top element turns out to be the maximum. This fact indicates that these keys convey the final result. Therefore the serial of keys can be regarded as the sequence that we want.

There are $n$-1 keys in this sequence. According to the statement that each key has two possible situations, there are $2^{n-1}$ different sequences in all. Every sequence has a probability of $(0.5)^{n-1}$ to be the final sequence. Use (1), we can obtain the initial information entropy.

$$H(x) = -\sum_{i=1}^{2^{n-1}} 0.5^{n-1} \log_2(0.5^{n-1})$$
$$= n - 1 \quad (4)$$

As $H_0(x)$ equals to 0, with the help of (2) the minimum computational complexity is got from the result of the calculation.

$$\text{Time complexity} = \Delta H(x) = n - 1 \qquad (5)$$

Therefore the calculated minimum computational complexity is O($n$). According to the survey of various algorithms, the computational complexity of the fastest program which is able to solve the problem is also O($n$). These two computational complexities are the same in such model. This indicates that the theory of the consistency between the information entropy change and the minimum computational complexity of the computer program is proved in this issue.

### 3.2. Modeling of Sorting a Group of Numbers

Being similar to the circumstances of seeking the maximum value, another type of sequence needs to be decided before we calculate the minimum computational complexity of sorting a group of numbers. Inspired by the idea of seeking the maximum value, keys between every two elements should be established in order to express the final relationships among data. Just as the circumstances in solving the problem, looking for the maximum, each key has two possible situations. If we decide the value of every key, the correct sequence will be presented. With the thinking as before, the sequence of these keys is considered what we want. Meanwhile, the number of the sequences is considered to be $2^{0.5n(n-1)}$ with the help of the permutation and combination theory. According to this method to do so, we will get a result like this:

$$H(x) = 0.5n(n-1) \qquad (6)$$

Therefore, the minimum computational complexity is O($n^2$). This computational complexity is the same with that of Bubble sort and insertion sort [10]. These methods are fast enough, however, just as we know, the fastest algorithm is merge sort the computational complexity of which is able to reach O($n\log_2(n)$) [10]. O($n\log_2(n)$) is much smaller than O($n^2$), this fact indicates that O($n^2$) is not the minimum computational complexity and O($n\log_2(n)$) takes place of it. As a result, there may be a contradiction in this theory. Nevertheless, if we consider this issue more carefully, we will discover that we have ignored some important point, the independence of the keys. For example, if there are only three numbers, all situations which are based on the theory of permutation and combination are shown in the following figures:

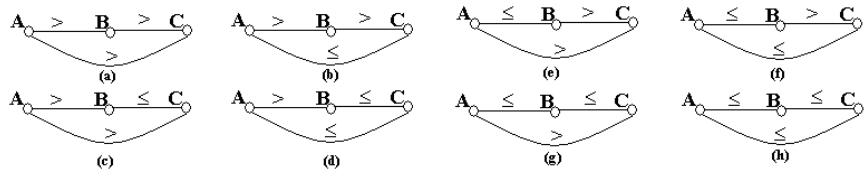

Figure 2. Situations of three elements

In reality, not all of the eight situations are able to occur. Fig. 2(b) and Fig. 2(g) are two examples of these circumstances. Taking account of the size relationships among them, there are contradictions in the two figures cited above. Thinking about Fig. 2(b) in details, we'll get the following information: A is greater than B; B is greater than C; A is not greater than C. According to the first two points, A should be greater than C, however Fig. 2(b) provides an opposite relation. Therefore the situation in Fig. 2(b) does not exist. The same conclusion is drawn from Fig. 2(g).

As the total number of data grows, more and more situations appear and we directly regarded them as legal ones, but in fact they should not be included in the collection of the sequences. If we exclude them, the total number of the situations will correspondingly reduce to $n!$. It is easier to think in another way in order to explain why the number equals to $n!$ after reduction. When the serial of keys are decided, one sequence of the numbers is determined. Meanwhile one sequence of the numbers indicates one sequence of "Greater" or "Not greater". This is true, because the sequence that we choose is able to express the characteristic of the final result and the sequence of the numbers is exactly the result in the problem of sorting a group of numbers. Therefore, the sequence of the numbers is capable of being the state $x_i$. Based on the permutation and combination theory again, the number of the states is $n!$. For the reason that every state has the same probability to be the final one, (1) and (2) provide us the following equation:

$$\text{Time complexity} = \log_2(n!) \qquad (7)$$

Computational complexity is an approximation of a program's speed. Moreover, the number of the elements should be large enough in order to make the approximation meaningful. In the case that $n$ is a large number, there is a classical approximation.

$$\ln(n!) = n\ln(n) - n \qquad (8)$$

Based on (7), we can get the following result:

$$\begin{aligned}\log_2(n!) &\approx n\log_2(n) - n\log_2(e) \\ &\approx n\log_2(n)\end{aligned} \qquad (9)$$

Therefore, the minimum computational complexity is $O(n\log_2(n))$. The result is the same with that of the merge sort [10]. This fact confirms the correctness of the theory again.

## 4. NOTICES OF MODELING PROBLEMS

Even though there is not a precise criterion of modeling problems, several notices should be clarified in order to get the correct result.

### 4.1. Relationship Between Final Results and the Sequence

In the second problem, we are able to choose the permutation of the numbers as the sequence because the final result is one of the permutations. However, such method can not be generalized.

Considering the problem of seeking the maximum value, if we regard the final result as the sequence, then an element is considered as the sequence and the number of the states will be $n$. According to (1), we are easy to find out that the minimum computational complexity is $\log_2(n)$ in this model. As we know, the computational complexity of solving this problem is not able to decline to this amount. Therefore the final result can not serve as the state in all problems.

However being aware of which problem's final result is capable of being the sequence is not enough, we need to discover the reason why some of them are able to be the sequence and the others are not.

In the problems of sorting a group of numbers, the final result is a serial of numbers in order. We regard this type of sequence as the state and we are able to calculate the minimum computational complexity which conforms to that of the fastest program. The reason why this consistency is established can be explained in the following way. The result of sorting a group of numbers is a

77

serial of numbers in order. The only feature of the result is the order. Therefore the final result itself is able to express its characteristic. We call this trait information independence. When the problem is solved, the information entropy reduces to 0. This means that the sequence we choose is decided. Fortunately, the sequence we determine is the final result and because of its information independence the final result contains all the final relationships among data. Thus, in this problem we can regard the final result as the sequence.

Nevertheless, the circumstances of seeking the maximum value are different from those of sorting a group of numbers. The final result of seeking the maximum value is the maximum number. Even though the maximum number shows us its magnitude, it is not able to convey its characteristic, being the maximum. Therefore, the consequence of this problem is not information independent and this fact leads to the invalidity of regarding the final result as the sequence.

In a nutshell, the sequence which is considered to be the state should express the final relationships among data. Only if the final result is information independent, is it possible to substitute the final result for the sequence in order to simplify the state calculation.

### 4.2. Final Information Entropy and the Sequence

In order to determine which sequence is able to be taken as the state, the imprecise criterion is not enough. For instance, in the problem of looking for the maximum value, besides the sequence we choose, a sequence that contains the relationships between each two numbers accords with the criterion too. Therefore, another factor should be taken into account.

Recalling the discussion in Section 3, when a problem is solved, the information entropy declines to 0. This means that the sequence is determined. However this fact will not happen, if we regard the sequence which expresses the relationships between each two numbers in the problem of seeking the maximum value. When the maximum value is located, there are only $n$-1 keys determined in the sequence and the rest of the keys remains unknown. Therefore, the entropy of information is not 0 in this case and the magnitude of the information entropy should be calculated again. This will bring in more work. Not only do we need to calculate both initial and final information entropies, we also need to determine the locations of the $n$-1 keys in the sequence.

Whereas, thing is not that easy as it looks like. The locations of the $n$-1 keys are able to provide several different states. Meanwhile, the sequence itself has several states too. It seems right to take these two types of states into account, but these two types of states are not totally different, there is an overlap between them. If we consider in this way, modeling this problem will be more complex and a great amount of extra work should be done.

As a result, it is a good way to model problems with the sequence that makes the final information entropy decline to zero. We call this type of sequence 0-type sequence. Choosing the 0-type sequence as the state is a routine way to model problems. However in some problems, the 0-type sequence is not easy to find. So we are able to regard some sequences that are in accord with the imprecise criterion as the state. If we are easier to handle the complex relationship between those two types of states than to establish the 0-type sequence, it will be a better way to select this sequence instead of the 0-type sequence.

### 4.3. Equation Simplification In Specific Circumstances

While calculating the computational complexity of the two problems in Section 3, we can find out that the equation is able to be simplified. Because the possibilities of all states are the same and the final information entropy is zero, the equation can be expressed in the following form (*n* is the number of the states):

$$\text{Time complexity} = \log_2(n) \tag{10}$$

This simplified equation discloses another characteristic of the states in the models of problems. That is the equality of the states' possibilities. This feature is a factor that we need to consider when we are going to select the sequence and it is also an inevitable result because the sequence satisfies the imprecise criterion. The necessity can be proved by the following evidence.

If some states' probabilities of being the final sequence are different from others, this will indicates that these states have different numbers of sub-states. Since each state is able to be the final sequence, meanwhile the sequence and the result is one to one, therefore if a state with one sub-state can be the final sequence and a state with two sub-states is also capable of being the final sequence, the format of the final result is not determined. However in a specific problem the format of the final result is settled. Thus, it is not possible to make a state with different probability.

## 5. NATURE OF SOLVING ISSUES WITH COMPUTER PROGRAMS

In Section 4, the equation to calculate the computational complexity is simplified. According to (10), there is a logarithmic relationship between the minimum computational complexity and the number of the states.

As everyone knows, the computational complexity of the fastest program for searching, bisearch also has a logarithmic relationship with the number of the elements. However this coincidence does not happen occasionally. In Section 3, solving problems using computer programs is interpreted as determining which state is able to be the final state. As the theory says, at the beginning of solving problems, there are a number of states and we don't know which state is capable of being the final one. These states act as a list and our target is to find out which state is in accord with the characteristics of the final result. The list is named as lookup table of problems and the final result is called searching target. Therefore, solving a problem with computer programs is transformed into a problem of searching. The nature of resolving issues turns out to be ravelling out a problem of searching.

Since the minimum computational complexity of searching is $\log_2(n)$ (*n* is the number of the elements), so the minimum computational complexity of any problem is able to reach $\log_2(n)$ (*n* is the number of the states).

## 6. REDEFINITION OF INFORMATION ENTROPY

In Section 5, solving a problem by computer programs can be considered as a searching problem. Hence the minimum computational complexity is related to the number of comparisons which are required to find the searching target. Based on the theory, if the connection between the minimum computational complexity and the information entropy is valid, there will be a relationship between information entropy and the corresponding searching problem. Considering the fact that information entropy has a similar logarithmic relationship to that of searching problems, a redefinition of the information entropy is proposed in the field of settling issues.

In the redefinition of information entropy, it is defined as the minimum number of comparisons, which computers need to execute in order to find the searching target in the lookup table of the problem in the worst case. This definition focuses on using computer programs to solve problems. Meanwhile it is also in accord with the original definition proposed by Shannon. Therefore the new definition is a special case of Shannon's definition.

Computer is a discrete binary system. It is able to resolve problems with definite results. According to the discussion in Section 4, the states of the problem are equiprobable. Thus, the equation of the information entropy can be reduced to the simple logarithmic form while considering such problems. This is the reason why we replace the original definition by the new one in this field. Moreover, this definition helps us to think about this type of issues more directly and conveniently.

Recalling the microcosmic definition of the thermodynamic entropy, we can find out that the new definition of the information entropy is similar to it. The only differences between them are the coefficient and the base number of the logarithm.

The coefficient of the thermodynamic entropy is $k$ ($k$ is Boltzmann constant) and 1 is the coefficient of the information entropy (new definition). This difference is caused by the differential definition of the thermodynamic entropy. In the differential definition, the thermodynamic entropy is given by the following equation:

$$dS = \frac{dQ}{T} \qquad (11)$$

As we know the information entropy has no unit, but the unit of the thermodynamic entropy is J/K. The unit of the thermodynamic entropy is given by this definition. Coincidentally, the unit of $k$ is also J/K, so if we modify the equation into the following form, these two entropies are closer to each other.

$$dS = \frac{dQ}{kT} \qquad (12)$$

After the modification of the thermodynamic entropy, the equations of these two entropies are still different in the perspective of the base number. This difference can be explained as the difference between the binary system and the natural system. Let's see the Taylor expansion of $e$ and 2.

$$e = 1 + 1 + \frac{1}{2!} + \frac{1}{3!} + ... + \frac{1}{n!} + ... \qquad (13)$$
$$2 = 1 + 1 \qquad (14)$$

The form of a term in the Taylor expansion of $e$ is like the possibility of an $n$-element-permutation (permutation of $n$ elements) to be the final permutation, if the $n$-element-permutations are equiprobable. Moreover, the binary system is able to resolve a zero-element-problem ($n$-element-problem means that the object for comparison is consist of $n$ elements) or a one-element-problem with only one comparison. If there is no element or only one element in the object, the possibility of getting the object we want will be 1 and if there are $n$ elements in the object, probability of obtaining the right object is the same with the ($n$+1)th term. Therefore the terms in the Taylor expansion can be explained as the possibility of achieving the right object and we also call the possibility the possibility of the $n$-element-problem.

In (14), the base number of the binary system is the sum of the two possibilities of zero-element-problem and one-element-problem. Meanwhile these two problems are able to be solved with only one comparison. However, $e$ is the sum of all the possibilities of $n$-element-problems. Therefore, a hypothesis that the natural system is able to solve an arbitrary $n$-element-problem with only one comparison is proposed and the system like the natural system is named as the infinite system.

In a nutshell, the thermodynamic entropy can be regarded as a special case of the generalized information entropy (the generalized information entropy means the information entropy that is able to be defined not only in binary system, but also in other systems, in other words, it means that the base number can be changed) [1].

## 7. CONCLUSION

As the result of the assumption that the process of solving a problem is the same as making a change in the information entropy and the postulation that an operation in computer program is able to generate 1 bit information, the theory that the minimum computational complexity of settling a specific issue with computer programs equals to the information entropy change in the model of the problem with special characteristics is established.

Seeking the maximum value and sorting a group of numbers are served as evidence to support this theory. The coincidence that these two computational complexities equal to the results of the calculation backs the validity of the principle. Meanwhile, (2) is proposed to calculate the minimum computational complexity.

While we are calculating the minimum computational complexity based on the information entropy, a pivotal problem of how to modeling issues arises. Thus, several notices are put forward in order to make the process of modeling problems more complete.

The first notice is about the imprecise criterion. This notice remind us that not all the problems are able to regard their final results as the sequences. If and only if the final result of the problem is information independent, is it correct to choose the final result as the sequence. In other words, the only criterion of selecting the sequence is that the sequence must be able to convey all the final relationships among data and taking the final result as the sequence is only a simplification of a special case.

Secondly, while selecting the sequence, there are a lot of sequences that are in accord with the imprecise criterion. This notice acts as a guide for us to choose the most befitting sequence among them. In the notice, the sequences are separated into two types. One is the sequence with zero final information entropy, which is called 0-type sequence; the other is the sequence with non-zero final information entropy. While calculating the number of the states, the circumstances of the sequence with non-zero final information entropy are very complex, but the cases of the 0-type sequence are easily figured out. However, the 0-type sequence is not usually established easily. If we can consider the complex circumstances of the sequence with non-zero final information entropy clearly, we will be able to take such sequence instead of the 0-type sequence. Therefore, the notice indicates that we should choose the 0-type sequence to make the calculation easier in general cases and we can select the sequence with non-zero final information entropy in some special cases, if we are able to think about the circumstances clearly.

An important characteristic of the states in the models of the problems is clarified in the final notice. This characteristic is described as the equiprobability of the states. In the analysis, the characteristic is proved to be necessary, because the result of the problem has a certain format. Meanwhile the equiprobability also leads to a simplification of the equation to calculate the minimum computational complexity.

After the simplification of the equation is made, the similarity between the calculated result and the minimum computational complexity of the searching problem is discovered. Such similarity arouses a further consideration of the relationship between the searching problem and the nature of solving problems with computer programs. Since the minimum computational complexity of searching has the same form with that of the simplified information entropy and there are lookup table and the searching target in the model of the problem, so the nature of solving problems with computer programs is regarded as resolving searching problems.

With the help of the discovery that the nature of solving issues with computer programs is no more than ravelling out searching problems, the information entropy is redefined as the minimum number of comparisons, which computers need to execute in order to find the searching target in the lookup table of the problem in the worst case. This redefinition is in accord with the original definition made by Shannon. Meanwhile a little modification is made to the thermodynamic entropy. This modification leads to a hypothesis that the thermodynamic entropy is a special case of the generalized information entropy which means the information entropy that is able to be defined not only in binary system, but also in other systems (in other words, it means that the base number can be changed).

In summary, all the conclusions above herald a new field of science to be developed.

## REFERENCES

[1]  Chen Gang, "Shannon Information Model in E-commerce Information Analysis," in International Joint Conference on Artificial Intelligence, pp.580-583 (2009)

[2]  Sergio Verdú, "Fifty Years of Shannon Theory," IEEE Transactions on Information Theory, vol. 44, no. 6, pp.2057-2078 (1998)

[3]  Yong WANG, "Analyses on Limitations of Information Theory," in International Conference on Artificial Intelligence and Computational Intelligence, pp.85-88 (2009)

[4]  Shannon, Claude Elwood, "A Mathematical Theory of Communication," Bell System Technical Journal, 27, pp.379-429, pp.623-656 (1948)

[5]  Shannon, Claude Elwood, "The Bandwagon," IEEE Transactions on Information Theory, 2, p.3 (1956)

[6]  A. J. Hatfield and K W. Hipel, "Understanding and Managing Uncertainty and Information," in IEEE International Conference on System, Man and Cybernetics, pp.1007-1012 (1999)

[7]  Ming Zhao, Xiulan Ye, Ke Han and Yun Li, "Research on Digital Image Edge Detection with Local Entropy and Fuzzy Entropy Algorithms," in IEEE International Conference on Information and Automation (ICIA), pp.2477-2482 (2010)

[8]  Xiang Xie, Xue-yun Zang and Zhong-liang Guan, "Analytic Demonstration on the Irrationality of Negative Entropy Principle," in International Conference on Management Science and Engineering, pp.109-115 (2008)

[9]  Xiaolong Wang, Yunjian Ge and Xiujun Wang, "Research on Manthematical Theory of Information Acquisition," in Proceedings of International Conference on Information Acquisition, pp.59-70 (2004)

[10]  Sardar Zafar Iqbal, Hina Gull and Abdul Wahab Muzaffar, "A New Friends Sort Algorithm," in 2nd IEEE International Conference on Computer Science and Information Technology, pp.326-329 (2009)

**Authors**

**Xue Wu** was born in Changchun, China. He is currently an undergraduate student from Tsinghua University. His major is Microelectronics. His interest is in mixed-signal circuit and RF circuit design. He is also interested in information theory which will be served as a subfield of him.

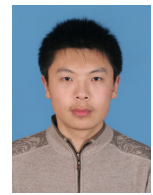